\documentclass[12pt,notitlepage,nofootinbib,superscriptaddress,showkeys]{revtex4-1}

\usepackage{graphicx}
\usepackage{psfrag}

\usepackage{amsmath,amssymb,mathrsfs}
\usepackage[bookmarks=false]{hyperref} 
\hypersetup{pdfstartview=FitH,pdfhighlight=/O,colorlinks=false}

\usepackage{cases}
\usepackage{subfigure}

\usepackage[left=1.2in,right=1.2in,top=1.5in,bottom=1.5in]{geometry} 
\usepackage{enumitem}

\begin{document}

\title{\large \bf Last gasp of a black hole:\\ unitary evaporation implies non-monotonic mass loss}

\author{Eugenio Bianchi}\email{ebianchi@gravity.psu.edu}
\affiliation{Institute for Gravitation and the Cosmos, Physics Department,
Penn State, University Park, PA 16802, USA}
\author{Matteo Smerlak}\email{msmerlak@perimeterinstitute.ca}
\affiliation{Perimeter Institute for Theoretical Physics, 31 Caroline St.~N., Waterloo ON N2L 2Y5, Canada}

\date{\small\today}

\begin{abstract}
\noindent {\bf Abstract}. We show within the usual two-dimensional approximation that unitarity and the restoration of Minkowski vacuum correlations at the end of black hole evaporation impose unexpected constraints on its mass loss rate: before disappearing the black hole emits one or more negative energy burst, leading to a  temporary increase of its mass.
\end{abstract}

\maketitle

\section{Introduction}
Quantum-mechanical effects allow black holes to radiate and therefore---by energy conservation---to lose their mass \cite{Hawking:1974sw}. The flux of radiation emitted by a large black hole of mass $M$ as computed by Hawking in 1974 is about one Hawking quantum with energy $\sim \hbar/M$ per light-crossing time $M$, that is a flux $F\sim \hbar/M^2$ (we use units $c=1$ and $G=1$). The rate of mass loss is therefore
\begin{equation}
\dot{M}= -F\sim -\frac{\hbar}{M^2}\,.
\label{eq:mass-loss-extimate}
\end{equation}
When extrapolated to late times, this result implies that the black hole will radiate at an increasingly high rate till its mass vanishes after a finite time $\tau\sim M_0^3/\hbar$, where $M_0$ is the initial mass of the black hole.

The radiation emitted by the black hole up to a time $t$ is in a mixed state with Planckian distribution at the Hawking temperature $T\sim \hbar/M$. The disappearance of the black hole at the end of the evaporation process poses therefore a puzzle, the puzzle of information loss: a pure state of collapsing matter can form a black hole that then evaporates leaving behind only the thermal Hawking radiation emitted. This phenomenon results in a net increase of the Von Neumann entropy of an isolated quantum system, in contradiction with the \emph{unitarity} of quantum mechanics.

Despite initial attempts to accommodate the loss of unitarity and predictability in gravitational phenomena  \cite{Hawking:1976ra}, it is today widely believed that the evaporation process should be ultimately unitary. A possible resolution of the paradox is that quantum effects not taken into account in the original Hawking calculation provide a resolution of the black hole singularity and a disappearance of the event horizon that protects it. In the absence of an event horizon, light can eventually escape to infinity and bring all the information needed for the restoration of unitarity. In this scenario a black hole is a metastable bound state of the gravitational field that traps light for a long but finite time within its apparent horizon.

We assume that the spacetime geometry of such unitary black holes can be described at all times by an effective classical metric with small quantum fluctuations\footnote{\label{foot:ell} It is immediate to estimate the size of the core of a black-hole-like object with the singularity resolved at a Planck-scale energy density: if we concentrate all the mass $M$ of the black hole in a core of size $\ell$ we find a density $\varrho\sim M/\ell^3$. A Planck scale upper bound $\varrho\sim \hbar^{-1}$ results in a core of size $\ell\sim (M\,\hbar)^{1/3}$. For a solar mass black hole this size is $10^{12}$ times larger that the Planck length $\ell_P=\hbar^{\,1/2}$ and can be described by a classical effective metric.} \cite{Hayward:2006fn,Rovelli:2014tm,Frolov:2014wc}. In this situation, the unitarity of Hawking evaporation can be phrased and studied using the methods of quantum field theory in curved spacetimes. Using these methods we show that the requirement of unitarity puts strong constraints on the form of the flux of the radiation emitted by the black hole. In particular we show that---to the extent that the usual two-dimensional reduction is a reliable guide---a unitary evaporation process is always accompanied by negative energy bursts reaching future null infinity. This phenomenon results in non-monotonic mass loss: an evaporating black hole temporarily increases its mass---``gasps''---before dying a unitary death.

\section{Ray-tracing and energy flux}
In an asymptotically flat spacetime, unitarity of the Hawking evaporation process can be formulated as an $S$-matrix problem. Consider a free massless quantum field. At past null infinity $\mathcal{I}^-$, the quantum field is prepared in the vacuum state $|0_-\rangle$. Propagation in a dynamical background results in a new state at future null infinity $\mathcal{I}^+$. The evolution from  $\mathcal{I}^-$ to $\mathcal{I}^+$ is unitary if the final state is a pure state in the out Fock space. In this case the final state can be written as $|\Psi\rangle=U|0_+\rangle$ where $|0_+\rangle$ is the vacuum state at $\mathcal{I}^+$ and $U$ a unitary operator. (Clearly if an event horizon is present, as in Hawking's original derivation, $\mathcal{I}^+$ is not a Cauchy surface for the massless field, the final state at future null infinity is mixed and unitarity between $\mathcal{I}^-$ and $\mathcal{I}^+$ is lost. Therefore, we only consider black hole spacetimes with $\mathcal{I}^-$ completely determined by data on $\mathcal{I}^+$.)

Let us focus on the spherically symmetric case. We restrict attention to its two-dimensional structure in the time-radial directions, and define an affine parameter $v$ on $\mathcal{I}^-$ labelling radially ingoing light rays. Similarly at $\mathcal{I}^+$ we define an affine parameter $u$ and use it to label radially outgoing light rays.\footnote{We partially fix the ambiguity in the choice of the affine parameter $u$ by demanding that the future-pointing null vectors $l=\partial_v$ and $n=\partial_u$ have scalar product $l\cdot n=-1$ at spatial infinity.} Infalling light rays that reflect off the center at $r=0$ define a one-to-one mapping between $\mathcal{I}^+$ and $\mathcal{I}^-$,
\begin{equation}
v=p(u)\,.
\end{equation}
The assumption that no event horizon forms corresponds to the \emph{ray-tracing} function $v=p(u)$ being onto. 

Remarkably, the ray-tracing function encodes all the relevant physics of the Hawking process \cite{Hawking:1974sw,Barcelo:2011cf}. In particular, the renormalized energy flux at $\mathcal{I}^{+}$, $F(u)=\langle \Psi|T_{uu}|\Psi\rangle - \langle 0_+|T_{uu}|0_+\rangle$, is simply given by \cite{Birrell:1982ix}
\begin{equation}\label{eq:riccati}
F(u)=\frac{\hbar}{48\pi}\big(k(u)^2+2\,\dot{k}(u)\big)\,,
\end{equation}
where the `peeling' function $k(u)$ is defined by
\begin{equation}
k(u)=-\frac{\ddot{p}(u)}{\dot{p}(u)}.
\end{equation}
A steady production of Hawking particles corresponds to a regime of approximately constant $k(u)$, i.e. to an adiabatic phase where $|{\dot{k}}/{k^2}|\ll 1$. In this regime the Hawking particles are distributed according to a Planck spectrum with temperature $T=\hbar\, k/2\pi$ and energy flux $F=\pi T^2/12\hbar$. For a collapse that results in a Schwarzschild black hole the peeling function approaches $k=1/4M$ and the standard result for the mass loss Eq. (\ref{eq:mass-loss-extimate}) is reproduced.

\section{Entanglement entropy and Page curve}
Unitarity imposes strong conditions on the form of the ray-tracing function $v=p(u)$ of a classical spacetime. The function must be invertible to avoid the presence of an event horizon, its derivative $\dot{p}(u)$ must go to a constant as $u\to \pm \infty$ to guarantee that the final state $|\Psi\rangle$ is normalizable, and its second derivative (or more specifically $k(u)$) must vanish as $u\to \pm \infty$ to ensure that the final state has finite energy \cite{Walker:1984vj,Ashtekar:2008jd}. Under these conditions the final state is pure. However, much more is generally required of the final state of black hole evaporation. If a black hole disappears in a finite (retarded) time $u$, all the correlations in the quantum field after this time $u$ are expected to coincide with the ones of a field in the Minkowski vacuum state. The entanglement entropy of the state $|\Psi\rangle$ at $\mathcal{I}^+$ provides a precise measure of these correlations.

We divide $\mathcal{I}^+$ in two semi-infinite intervals, $(-\infty,u)$ and $(u,+\infty)$. 
The entanglement entropy of field degrees of freedom between the two parts is defined as
\begin{equation}
\mathcal{S}_{|\Psi\rangle}(u)=-\text{Tr}_{(-\infty,u)}\,\big(\rho(u) \log \rho(u)\big)
\end{equation}
where $\rho(u)=\text{Tr}_{(u,+\infty)}|\Psi\rangle\langle\Psi|$ is the reduced density matrix \cite{Bombelli:1986rw}. This quantity is in general divergent. Similarly to what is done for the flux we introduce a UV regulator and compute the renormalized entanglement entropy defined by the difference $S(u)=\mathcal{S}_{|\Psi\rangle}(u)-\mathcal{S}_{|0_+\rangle}(u)\,$. This difference is finite when the regulator is removed. Using the two-dimensional methods introduced in \cite{Holzhey:1994kc} we recently found that $S(u)$ is related to the peeling function by the remarkably simple identity \cite{Bianchi:2014qua,Bianchi:66YvM2Kp}:
\begin{equation}
S(u)=\frac{1}{12}\int_{-\infty}^u k(u')\,du'\,.
\label{eq:DS}
\end{equation}
At early times, before the evaporation begins, the entanglement entropy matches the one of the Minkowski vacuum state,
\begin{equation}
\lim_{u\to -\infty} S(u)\;=\;0\,.
\end{equation}
In the adiabatic phase of Hawking evaporation the entanglement entropy $S(u)$ grows linearly and matches exactly the statistical entropy of a gas of thermal quanta at the Hawking temperature. The purity of the state $|\Psi\rangle$, however, does not allow the entanglement entropy to grow indefinitely. In particular the requirement that after the disappearance of the black hole all the quantum correlations are restored to their Minkowski value corresponds to the condition\footnote{In terms of the ray-tracing function this condition corresponds to $\dot{p}(u)\to 1$ smoothly as $u\to \pm \infty$, i.e. no redshift or blueshift at initial and late times.}
\begin{equation}
\lim_{u\to +\infty} S(u)\;=\;0\,.
\end{equation}
A smooth $S(u)$ that starts at zero and returns to zero is often referred to as a ``Page curve'' \cite{Page:1993bv}. We refer the reader to \cite{Bianchi:66YvM2Kp} for explicit computations of $S(u)$ in two-dimensional spacetimes of interest.

\section{Black hole's last gasp}
The main question addressed in this paper is: What conditions do unitarity and the restoration of Minkowski correlations impose on the energy flux $F(u)$ emitted by the black hole during its life? The physical requirement of energy conservation 
\begin{equation}
\dot{M}(u)=-F(u)\,,
\label{eq:energy-conservation}
\end{equation}
translates these conditions on $F(u)$ into constraints on the mass law $M(u)$ of the black hole.

\begin{figure}[t]
\psfrag{a}{\hspace{-1.7em}$\frac{S(u)}{S_{\text{max}}}$}
\psfrag{b}{$u/\tau$}
\psfrag{c}{\hspace{-.3em}$\frac{M(u)}{M_0}$}
\psfrag{d}{$u/\tau$}
\includegraphics[height = 13.3em]{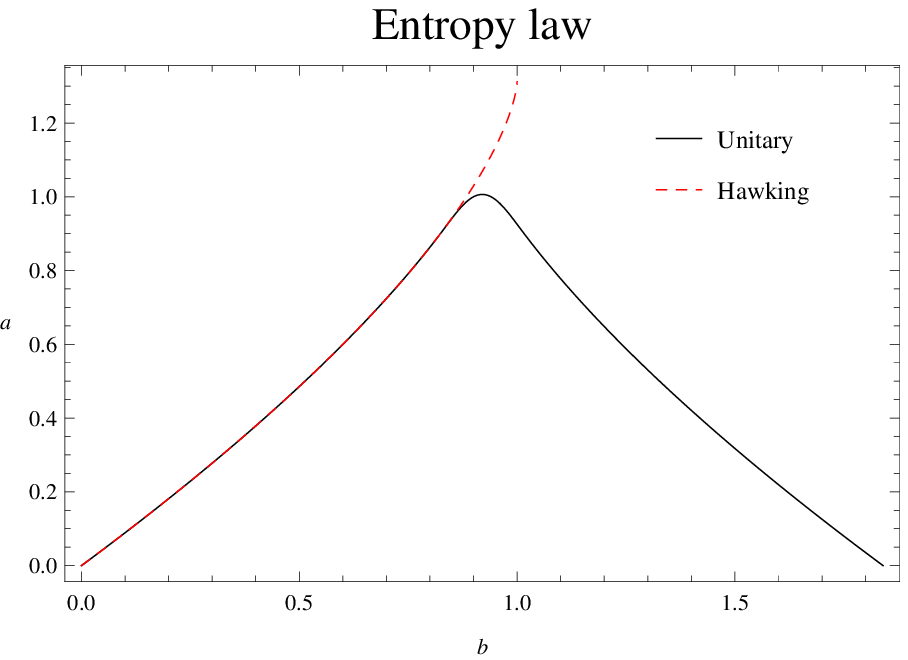}
\;\raisebox{2.7cm}{$\implies$} \!\!\!
\includegraphics[height = 13.3em]{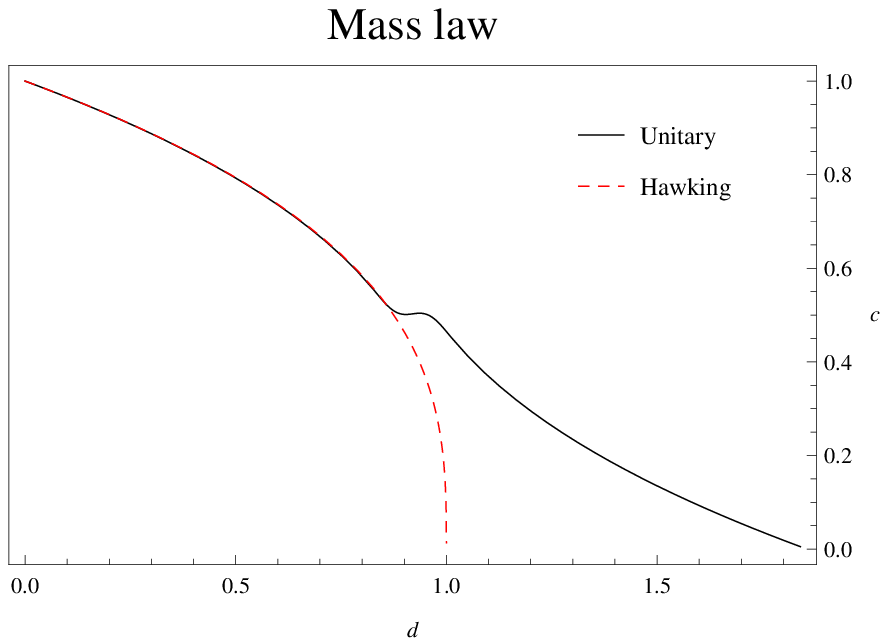}
\caption{Left: Example of a possible Page curve for unitary black hole evaporation. Right: Mass law derived from this Page curve; when the ``purification'' phase starts the mass of the black increases temporarily. The Hawking thermal entropy and mass law (dashed line) are for reference, and $\tau\sim M_{0}^{3}/\hbar$ denotes the Hawking evaporation time.}
\label{fig:gasp}
\end{figure}

The point of view adopted here is that the energy flux $F(u)$ is a given function (for instance because it has been measured) and the peeling function $k(u)$ is obtained by solving the differential equation \eqref{eq:riccati}. The question is if this flux is compatible with unitarity, i.e. with a peeling function $k(u)$ that vanishes smoothly at infinity. Introducing the ansatz $k(u)=2\,\dot{\psi}(u)/\psi(u)$, the Riccati equation (\ref{eq:riccati}) reduces to an ordinary differential equation
\begin{equation}
-\ddot{\psi}(u)+V(u)\psi(u)=0\,,
\end{equation}
the Schr\"odinger equation for a zero-energy resonance in the potential 
\begin{equation}
V(u)=\frac{12\pi}{\hbar}\,F(u).
\end{equation}
The four boundary conditions
\begin{equation}
\psi(u)\to 1\;,\qquad \dot{\psi}(u)\to 0 \qquad \text{for}\qquad u\to\pm \infty\,.
\label{eq:bc}
\end{equation}
correspond to perfect transmission and over-constrain the potential (Fig.\ref{fig:gasp}). In particular integrating the differential equation with the boundary conditions (\ref{eq:bc}),  we find the necessary condition
\begin{equation}
\int_{-\infty}^{+\infty}V(u)\, \psi(u)\, du\,=\,0\,.
\end{equation}

Now, using the relations (\ref{eq:DS}) and (\ref{eq:energy-conservation}) we can write this expression in terms of the renormalized entanglement entropy
\begin{equation}
S(u)=\frac{1}{6}\log \psi(u)
\end{equation}
and the mass loss of the black hole $\dot{M}(u)$, yielding
\begin{equation}
\int_{-\infty}^{+\infty} \dot{M}(u)\,\exp[6\, S(u)]\,du\,=\,0\,.
\label{eq:identity}
\end{equation}
This identity is the main result of this paper. In particular for the integral (\ref{eq:identity}) to vanish the rate of mass loss $\dot{M}(u)$ must change sign at least once: in a unitary evaporation process the mass of the black hole cannot decrease monotonically.


\providecommand{\href}[2]{#2}\begingroup\raggedright\endgroup

\end{document}